%% file: main.tex
\documentclass[12pt]{iopart}
\usepackage{comment,mfirstuc}
\usepackage{graphicx}% Include figure files
\usepackage{grffile}
\usepackage{algorithm}% http://ctan.org/pkg/algorithm
\usepackage{algpseudocode}% http://ctan.org/pkg/algorithmicx
\usepackage{hyperref}% add hypertext capabilities
\usepackage{soul}
\usepackage{listings}
\usepackage{multirow}
\usepackage{dcolumn}% Align table columns on decimal point
\usepackage{bm}% bold math
\usepackage{placeins}
\expandafter\let\csname equation*\endcsname\relax
\expandafter\let\csname endequation*\endcsname\relax
\usepackage{amsmath}
\usepackage{amssymb}
\usepackage{mathtools}
\usepackage{subfig}
\usepackage[bottom=1.0in]{geometry}
\usepackage{cite}
\usepackage{lineno}
\RequirePackage{xcolor}
\usepackage[misc]{ifsym}

\usepackage{tcolorbox} % Create coloured boxes (e.g. the one for the key-words)
\usepackage{stfloats} % Correct position of the tables

\usepackage{siunitx}

\usepackage{xspace}
\usepackage{colortbl}
\usepackage{array}
\newcolumntype{P}[1]{>{\centering\arraybackslash}p{#1}}
\newcolumntype{M}[1]{>{\centering\arraybackslash}m{#1}}
\hypersetup{ 
    pdfnewwindow=true,      % links in new window
    colorlinks=true,       % false: boxed links; true: colored links
    linkcolor=blue,         % color of internal links
    citecolor=blue,        % color of links to bibliography
    filecolor=blue,      % color of file links
    urlcolor=blue        % color of external links
}  

\algblock{Input}{EndInput}
\algnotext{EndInput}
\algblock{Output}{EndOutput}
\algnotext{EndOutput}

\def\be{\begin{eqnarray} &&} 
 
\def\ee{\end{eqnarray}}

\makeatletter
\newcommand{\mainmatter}{%
  \setcounter{footnote}{0}%
  \patchcmd{\@makefntext}{\fnsymbol}{\arabic}{}{}%
  \patchcmd{\@thefnmark}{\fnsymbol}{\arabic}{}{}%
  \def\@makefnmark{\textsuperscript{\arabic{footnote}}}
  \long\def\@makefntext##1{\parindent 1em\noindent
        \hb@xt@1.8em{%
            \hss\@textsuperscript{\normalfont\@thefnmark}}##1}%
}
\makeatother

\newcommand{\addComment}[2]{
  \expandafter\newcommand\csname #1\endcsname[1]{{\bf \color{#2} \capitalisewords{#1}:\,##1}}
  \expandafter\newcommand\csname #1cor\endcsname[2]{{\color{#2} \capitalisewords{#1}:\,\st{##1}{\bf ##2}}}
  \expandafter\newcommand\csname #1color\endcsname{#2}
}
\addComment{cris}{blue} 
\addComment{james}{red}

\newcommand{\epic}{ePIC\xspace} 
\newcommand{\gluex}{\textsc{GlueX}\xspace} 
\newcommand{\geant}{\textsc{Geant4}\xspace}

\begin{document}

% Keywords command
\providecommand{\keywords}[1]
{
  \small
  \textbf{Keywords:}  {\color{blue}#1
  }
}

\title[\scriptsize{Deep(er)RICH: Deep(er) Reconstruction of Imaging CHerenkov Detectors}]{Deep(er) Reconstruction of Imaging Cherenkov Detectors with Swin Transformers and Normalizing Flow Models} %One-Class Classification and 

%Diversity Identification
%\cris{One-Class Classification}}

%A Conditional Generative Approach to Anomaly Detection
%{Flux \& Mutability: A Conditional Generative Approach to Anomaly Detection \cris

%\author{F. Last$^{1^{\ast}}$} %\S, %x$^{5,\star}$, Z. Papandreou$^{5,\ddagger}$}

\author{C. Fanelli$^{1,2,\S}$, J. Giroux$^{1,\star}$, J. Stevens$^{2,\ddagger}$} %\S, %x$^{5,\star}$, Z. Papandreou$^{5,\ddagger}$}

\address{
$^{1}$ William \& Mary, Department of Data Science, Williamsburg, VA 23185, USA\\
$^{2}$ William \& Mary, Department of Physics, Williamsburg, VA 23185, USA\\
}

\ead{{\color{blue}
$^{\S}$cfanelli@wm.edu,
$^{\star}$ jgiroux@wm.edu, 
$^{\ddagger}$ jrstevens01@wm.edu 
}}

\vspace{10pt}
\begin{indented}
\item[]\today
\end{indented}

%\linenumbers

\begin{abstract}

Imaging Cherenkov detectors are crucial for particle identification (PID) in nuclear and particle physics experiments. Fast reconstruction algorithms are essential for near real-time alignment, calibration, data quality control, and efficient analysis. At the future Electron-Ion Collider (EIC), the ePIC detector will feature a dual Ring Imaging Cherenkov (dual-RICH) detector in the hadron direction, a Detector of Internally Reflected Cherenkov (DIRC) in the barrel, and a proximity focus RICH in the electron direction. This paper focuses on the DIRC detector, which presents complex hit patterns and is also used for PID of pions and kaons in the \gluex experiment at JLab.
We present Deep(er)RICH, an extension of the seminal DeepRICH work, offering improved and faster PID compared to traditional methods and, for the first time, fast and accurate simulation. This advancement addresses a major bottleneck in Cherenkov detector simulations involving photon tracking through complex optical elements.
Our results leverage advancements in Vision Transformers, specifically hierarchical Swin Transformer and normalizing flows. These methods enable direct learning from real data and the reconstruction of complex topologies.
We conclude by discussing the implications and future extensions of this work, which can offer capabilities for PID for multiple cutting-edge experiments like the future EIC.

%{\color{blue}Keywords: \textbf{conditional masked autoregressive flow, clustering, anomaly detection, one-class, neutral showers, jets}}

\end{abstract}


%\maketitle

\mainmatter

\input{1_introduction}

\input{2_data}
\input{3_architecture}

\input{4_analysis}

\input{5_impacts}

%\clearpage

\input{6_summary}

\section*{Code Availability}
The code is publicly available at \href{https://github.com/wmdataphys/DeeperRICH}{https://github.com/wmdataphys/DeeperRICH}.

%\clearpage

\section*{Acknowledgments}

% this is alphabetical reflecting author list
%
We thank William \& Mary for supporting the work of CF and JG through CF's start-up funding.
The authors acknowledge  William  \&  Mary  Research  Computing for providing computational resources and technical support that have contributed to the results reported within this article.
The authors acknowledge the \gluex collaboration for allowing us to use the simulation datasets described in this article.

%\clearpage
\section*{References}
\bibliographystyle{iopart-num}
\bibliography{biblio}

%\clearpage
%\input{Appendix}

\end{document}

%% file: 1_introduction.tex
\section{Introduction}\label{sec:intro}

Cherenkov detectors are extensively used in modern nuclear and particle physics experiments for charged particle ($\pi$, $K$, $p$) identification (PID). 
Cherenkov radiation is emitted in a cone shape along the particle's momentum direction when a charged particle moves through a dielectric medium at a speed greater than the phase velocity of light in that medium. The Cherenkov angle, $\cos(\theta_c) = \frac{1}{n\beta}$, depends on the particle's momentum ($ \propto \beta$) and the refractive index of the material ($n$). As momentum increases, the Cherenkov angles of different particles become similar, necessitating advanced PID methods such as deep learning \cite{fanelli2020machine}.
A Cherenkov detector typically uses single-photon detectors to identify charged particles based on the shapes of the detected hit patterns, which vary according to the particle's kinematics. Imaging Cherenkov detectors usually detect sparse, noisy rings, with the photon yield dependent on the particle's kinematics.

\begin{figure}[b]
    \centering
    \includegraphics[width=0.9\textwidth]{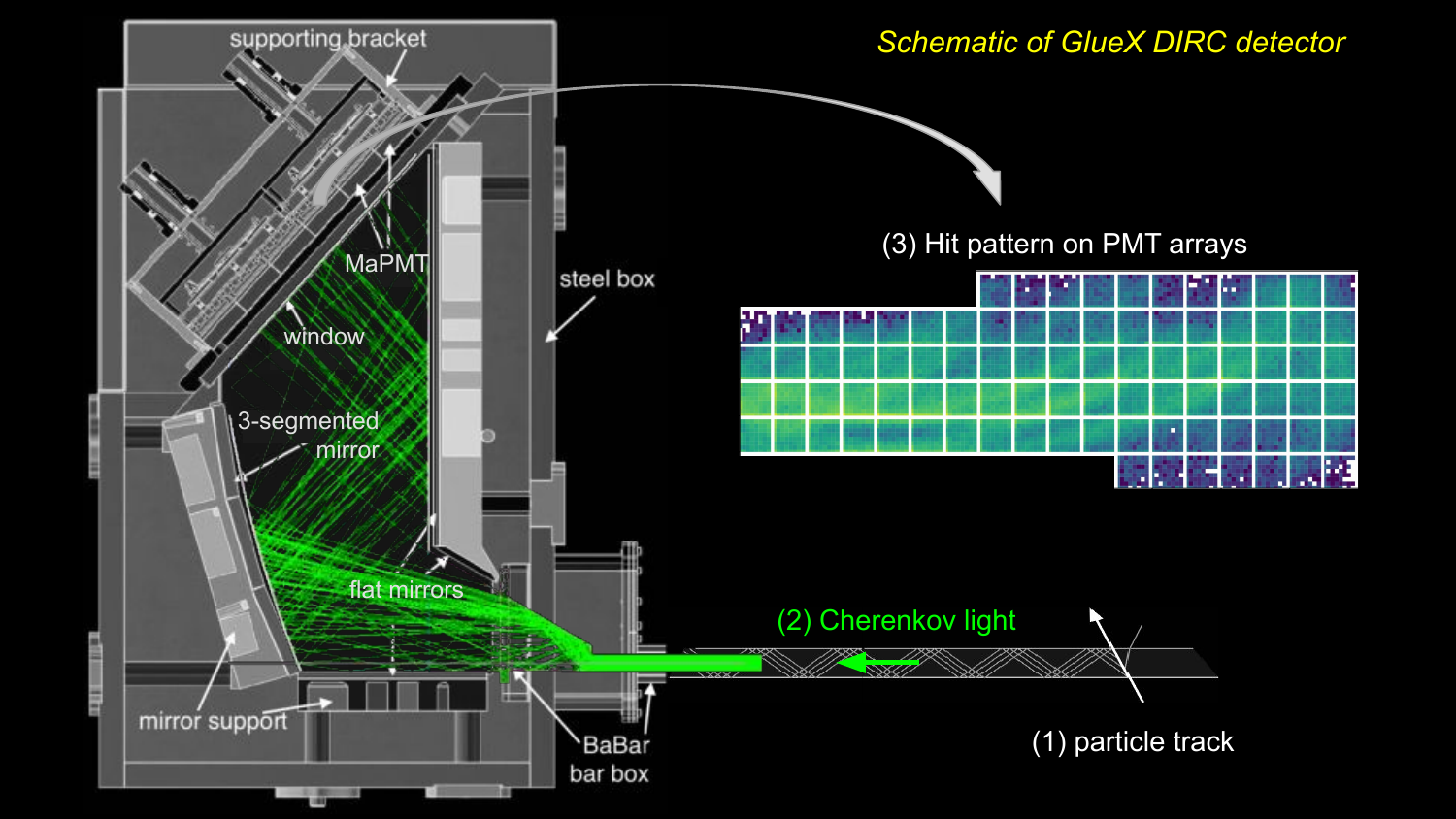}
    \caption{\textbf{Schematic of \gluex DIRC geometry:} A charged particle traverses the fused silica bar (step 1), generating Cherenkov light (step 2). This light undergoes internal reflection, reaching the optical box. Through a series of mirrors, the light is directed to an array of PMTs (step 3) for detection. The resulting hit pattern, which depends on the particle's kinematics, is illustrated in the final step, showing the accumulated hit patterns from multiple tracks with the same kinematics.}   
    \label{fig:dirc_geometry}
\end{figure}

Cherenkov detectors are central to the PID capabilities of experiments such as the future \epic experiment at the Electron-Ion Collider (EIC) \cite{khalek2022science} and \gluex at JLab \cite{adhikari2021gluex}, making them critical to the success of these research programs.
% JRS note: this sounds like propoganda without backing it up with evidence
%The \epic detector at the EIC will be a next-generation experiment aimed at unraveling fundamental questions about strong interactions in the universe. 
Cherenkov detectors will form the backbone of PID for charged hadrons in \epic, including a dual-RICH detector in the hadronic endcap \cite{cisbani2020ai}, a DIRC detector in the barrel region \cite{kalicy2024high}, and a proximity focus RICH in the electron region \cite{bhattacharya2023simulation}.
The DIRC detector exhibits more complex patterns, as light is contained by total internal reflection within a solid fused silica radiator \cite{adam2005dirc}. This preserves the angular information until the light reaches segmented photon sensors, resulting in intricate hit patterns.
This work focuses on the hit patterns observed in the DIRC detector currently operating in \gluex at JLab \cite{stevens2016gluex, patsyuk2018status} (see also Fig. \ref{fig:dirc_geometry}).
The data for this study, produced with \gluex software, correspond to full \geant \cite{GEANT4:2002} simulations of charged pions ($\pi$) and kaons ($K$) detected with the DIRC, challenging due to the proximity of their Cherenkov angles above 3 GeV/c. The DIRC detector in \gluex is located in the forward region, providing $3\sigma$ separation between $\pi$ and $K$ up to 3.7 GeV/c momentum.
It consists of 48 fused silica bars, segmented into 4 bar boxes, and two readout zones (optical boxes). The optical boxes, filled with distilled water and equipped with highly reflective mirrors, detect Cherenkov light through arrays of Multi-anode Photomultiplier Tubes (PMTs). Each PMT has 64 sensors arranged in an $8\times 8$ grid, with the PMTs themselves organized into a $6 \times 18$ array, providing the spatial location of the hit on the PMT plane (x, y) and time of arrival for each Cherenkov photon. A schematic of the \gluex DIRC geometry is provided in the following section.

The proposed research has three main goals: enhancing the distinguishing power of particles detected with Cherenkov detectors, developing faster and more accurate simulations essential for reconstruction, alignment, and calibration, and ensuring compatibility with near real-time and online applications through competitive inference times and portable solutions.
To achieve the goals of this research, we developed Deep(er)RICH, an extension of the pioneering DeepRICH architecture \cite{fanelli2020deeprich}, which was a first convolutional approach to Cherenkov PID in Nuclear Physics using hit-level information, coupling a Convolutional Neural Network (CNN) to the latent space of a Variational Autoencoder (VAE). It provided fast and accurate classification of pions and kaons at GlueX using hit-level information. 
Building on DeepRICH, our architecture uses hit-level information to form image representations of hits within the readout system of the GlueX DIRC, achieving for the first time full generalization over the phase-space of the charged particles through conditional learning and the power of hierarchical Swin Vision Transformers \cite{Liu_2021}. 

Our results on PID are compared to the established geometrical reconstruction, a method based on a look-up-table (LUT) that involves generating and tracking many photons at various angles as they exit each bar and travel to the PMT plane. For each hit $\vec{x} \equiv (x,y,t)$ recorded in a PMT pixel, the LUT provides possible photon-propagation vectors that could have resulted in the photon hitting that pixel. Particle identification is then achieved by computing the likelihood difference between $\pi$ and $K$ mass hypotheses utilizing the particle’s trajectory as measured by the tracking system. Since the exact exit point of each photon is unknown during LUT generation, photons are assumed to exit from the center of the bar face, resulting in per-bar images rather than per-particle images. We demonstrate how we outperform this traditional method over the whole phase-space of the charged particles.

Fast simulations must be faster than those simulated with \geant, addressing a major bottleneck in Cherenkov detector simulations involving photon tracking through complex optical elements. 
Previous research on fast simulation \cite{derkach2020cherenkov} bypasses low-level details and learns the output of the FastDIRC \cite{hardin2016fastdirc} simulation, which in turn uses the billiard method to project photons directly onto the readout plane based on the  reflections in the radiator and optical box. However, this approach is limited by its dependence on a simulation method to accurately reproduce real data.
Our approach on the other hand deeply learns the detector response from data to directly reproduce the observed hit patterns, using Normalizing Flow parameterized by several conditional Affine Coupling transformations \cite{Affine}.
These transformations encode the unknown probability distribution into a Multivariate Gaussian, conditioned on the kinematics of an incoming charged track through a neural embedding function \cite{freia,nflows}. 
Specifically, given a set of kinematics, we produce the mean and covariance matrices for the conditional Multivariate Gaussian through a neural network. 
This allows analytical calculation of likelihoods under an arbitrary Probability Density Function (PDF), and efficient sampling of hits from said PDF. 

As described in the following sections, our method achieves state-of-the-art performance in particle identification, produces accurate and fast simulations, and offers leading inference times for both PID and fast simulations.
The developed approach can be deployed on real data, given high-purity samples of $\pi$ and $K$ for PID, and can rapidly simulate complex hit patterns observed in real data with high fidelity.

%% file: 2_data.tex
\section{Data Preparation}\label{sec:data}

\begin{comment}
In this section we provide a description of what the data looks like and how dat have been preprocessed.

We work with hit patterns of Cherenkov photons detected by an electronic readout made by arrays of PMTs \cite{stevens2016gluex}. Each hit pattern is produced by an individual charged track at a given kinematics.

Fig. \ref{fig:sparse_hits} shows the sparse hit pattern produced by an individual track (hits colored ref), in comparison to the emergence of the true probability density function (PDF), \textit{i.e.}, the full hit pattern, when integrated over multiple tracks at the same kinematics. 
\end{comment}

In this section, we provide a description of the data and the preprocessing steps.

\begin{figure}[H]
    \centering 
    \includegraphics[width=0.85\textwidth,trim= 0.03cm 10.2cm 0cm 10.2cm, clip]{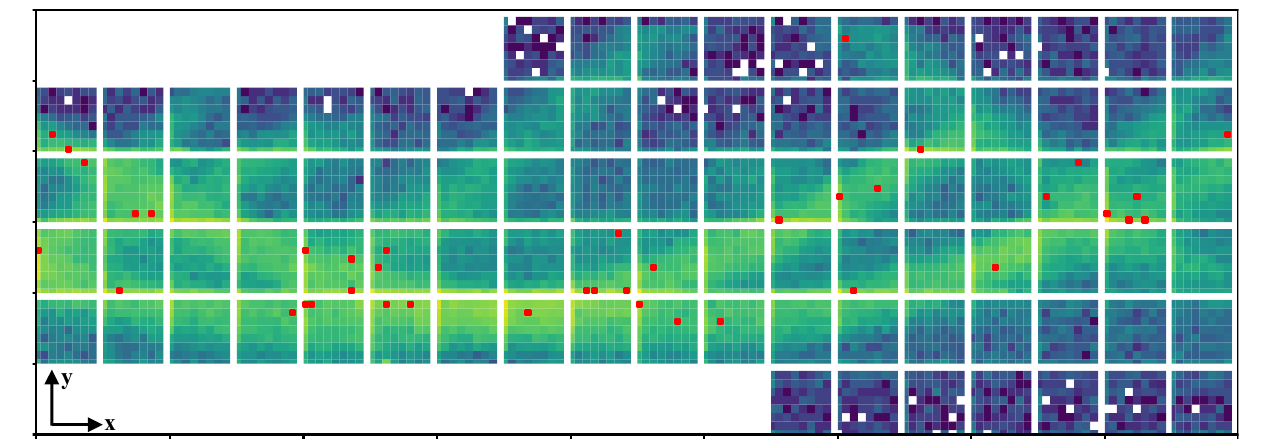} \\
    \caption{\textbf{Optical box output:} Individual tracks leave sparse hit patterns (red points) integrated over time on the DIRC readout plane, proving to be a challenge for convolutional-based networks to deal with. 
    %The hit pattern at the track level is approximately stochastic, in comparison to a 
    The denser hit pattern is obtained by accumulating multiple tracks with same kinematics. White zones are locations at which PMTs are not installed due to low accumulation of hits. %The white regions are not equipped with PMTs in the \gluex DIRC because they are areas with a low density of hits.
    }
    \label{fig:sparse_hits}
\end{figure}

We work with hit patterns of Cherenkov photons detected by an electronic readout consisting of arrays of PMTs \cite{stevens2016gluex}. Each hit pattern corresponds to an individual charged track with specific kinematics. Figure \ref{fig:sparse_hits} shows the sparse hit pattern produced by an individual track (hits colored in red), compared to the emergence of the true PDF, \textit{i.e.}, the full hit pattern, when integrated over multiple tracks with the same kinematics. Note that the white regions are not equipped with PMTs in the \gluex DIRC because they are areas with a low density of hits.

\paragraph{General Processing} 

In the case of \textit{Particle Gun} samples (\textit{i.e.}, simulation of individual tracks with given kinematics), processing data is relatively simple due to the single-track nature of the simulation. We can associate singular, identified charged tracks (measured by drift chambers) with the corresponding Cherenkov photon hits in one of the two optical boxes. The resulting output is stored at the track level, where each track has associated kinematics, hits, and other useful metadata for plotting purposes (such as the bar number and the impingement location on the bar).
Particle gun simulations allow full control over the generated phase-space, in which we generate charged pions and kaons approximately uniformly over the acceptance of the DIRC, corresponding to the ranges $0.5<|\vec{p}|<6.5$~GeV/c, $0<\theta<11^\circ$ and $0<\phi<360^\circ$. For this data, we require the momentum of the reconstructed track to be within $1 - 6.5$~GeV/c. 
%$\theta$ and $\phi$ are taken as the full region of coverage of the DIRC, corresponding to a width of approximately \SI{11}{\degree} in $\theta$ and \SI{360}{\degree} in $\phi$.

 \paragraph{Translation to sensor coordinate system}

The data structure produced by the DIRC corresponds to a PMT and pixel index for each individual hit in the detector array. The detector array is represented by a grid of $6 \times 18 $ PMTs, each of which consists of an $8 \times 8$ grid of `pixels'. These pixels are $6 \times 6\, mm$ photo-diodes which can be translated into an overall grid representation of $48 \times 144$ through Eq. \ref{eq:rowcol_transformation}, where $M_{PMT.}$ and $N_{pixel.}$ correspond to the PMT and pixel indices, respectively:\footnote{The $\lfloor \ \rfloor$ term denotes a floor division.}
\begin{equation}\label{eq:rowcol_transformation}
    D_{i,j} =
    \begin{cases}
        \begin{aligned}
        & \lfloor M_{PMT.}/18 \rfloor \cdot 8 + \lfloor N_{pixel.} / 8 \rfloor \\
        & (M_{PMT.} \mod 18) \cdot 8 + (N_{pixel.} \mod 8) 
        \end{aligned}
    \end{cases}
\end{equation}
Working in this coordinate system creates issues for NF models, given the fact that they are designed to work on continuous spaces. The above transformation results in a set of discrete values, representing a discrete-valued probability distribution over the sensor array. To overcome this, we first transform to the actual $x,y$ coordinate system of the sensor through Eq. \ref{eq:xy_transformation}, where the $6 \, mm$ and $2 \, mm$ shifts correspond to the physical dimensions of the pixels and spacing of the PMTs, respectively. The addition of $3 \, mm$ corresponds to the shift of the mapping to the center of the pixel.

%             x = col * 6. + (pmtID % 18) * 2. + 3.
%             y = row * 6. + (pmtID // 18) * 2. + 3.

\begin{equation}\label{eq:xy_transformation}
        \begin{aligned}
        & x = D_j \cdot 6 \, mm + (M_{PMT.} \mod 18) \cdot 2 \, mm + 3 \, mm \\
        & y = D_i \cdot 6 \, mm + \lfloor M_{PMT.} / 18 \rfloor \cdot 2 \, mm + 3 \, mm
        \end{aligned}
\end{equation}

Given the representation above, we are able to account for the issue of discrete values through a smearing operation. For a given hit on a pixel in the $x,y$ coordinate system, the probability of an individual photon striking anywhere on the face is approximately uniform. Therefore, we can apply a smearing operation without a loss of generality. To be specific, we introduce random uniform noise corresponding to the physical sensor limitations in $x$ and $y$ \textit{i.e.}, $\text{Uniform}(-3,3)$. Note that this is only done for training and the evaluation is done on the central representations. While it is possible to directly use the discrete values, computationally the models tend to become unstable during training resulting in large fluctuations.

 \paragraph{Image Formation}

For the Transformer, we form images using the coordinate system described by Eq. \ref{eq:rowcol_transformation}. In order to account for timing information, we also introduce a second channel of the same size which is populated with timing information from each individual hit. The resulting input to the network is of shape ($48,144,2$), where the second channel is scaled on the interval ($0,1$). 

%% file: 3_architecture.tex
\section{Architecture}\label{sec:architecture} 

%%%%%%%%%%%%%%%%%%%%%%%%%%%%%%%%%%%%%%%%%%%%%%%%%%%%%%%%%%
\subsection*{Vision Transformers}

Vision Transformers are powerful computer vision tools used to extract embeddings from patches of images through mechanisms such as Multi-Head Self Attention (MHSA). At a high level, we can envision this process as breaking the image into a series of patches, flattening these patches into embeddings (vectors), and then finding similarities between all embeddings through a series of dot products. This is done by populating three matrices, namely Query ($Q$), Key ($K$) and Value ($V$), all of which are parameterized through learnable transformation $\boldsymbol{x} W_{Q,K,V}$. In the case of MHSA, we can perform $n$ attention computations in parallel and independently of one another. The learned attention mask then takes the form of Eq. \ref{eq:MHSA}, where $i$ denotes the number of heads (parallel computations). 

\begin{equation}\label{eq:MHSA}
    \begin{aligned}
        \text{MHSA}(Q,K,V) & = \text{Concat}(H_1, ... \, , H_n) \cdot W_0 \\
        H_i & = \text{Softmax}(\frac{Q_i K_i^T}{\sqrt{d_K}}) \, V_i^T
    \end{aligned}
\end{equation}

\begin{figure}
    \centering
    \includegraphics[width=\textwidth]{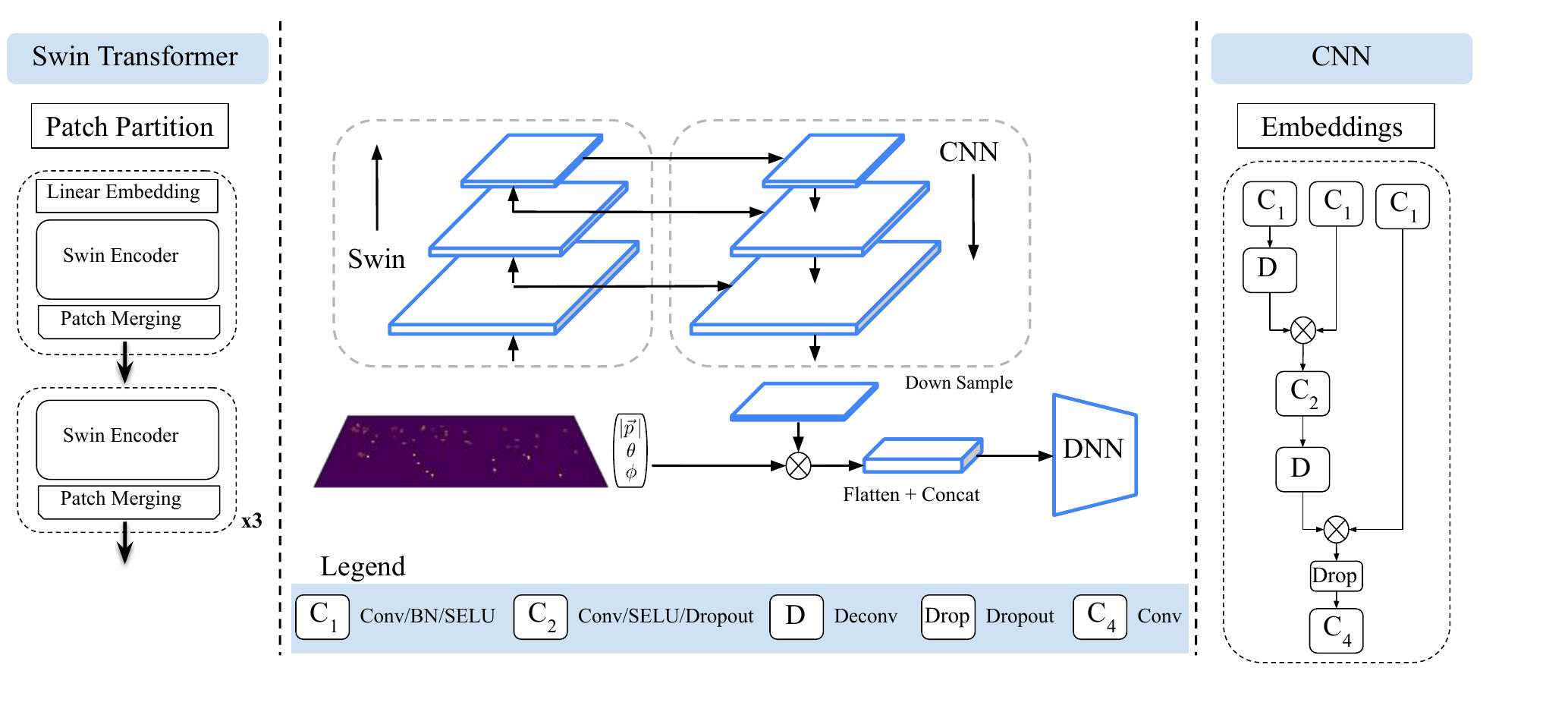}
    \caption{\textbf{Architecture flow chart:} Images of the optical box are processed through four consecutive Swin encoder blocks, producing feature maps at different resolutions. These outputs are fed in parallel to a Convolution Neural Network for recombination and downsampling prior to a flattening operation. The flattened vector is concatenated to the track kinematics and processed by a simple Deep Neural Network producing a binary label. }
    \label{fig:architecture}
\end{figure}

In the sparse image representations of the Optical Boxes (Fig. \ref{fig:sparse_hits}, overlayed red points), it is possible to have features existing on different resolutions, \textit{i.e.}, regions of higher and lower density. As such, we utilize Swin Vision transformers that emulate the structure of Feature Pyramid Networks, producing patched embeddings at different resolutions. \cite{Liu_2021,lin2017feature} These embeddings are fed to a convolutional-based network for recombination before finally being subject to a Deep Neural Network (DNN) for classification. Given that the Cherenkov angle depends strictly on the velocity, we want the network to learn as a function of the reconstructed momentum. Moreover, we want to characterize the location of the hit on the DIRC plane through the variables $\theta$ and $\phi$. As such, we also inject the kinematics (the momentum $| \vec{p} |$, the angle $\theta$ and the angle $\phi$) through a concatenation operation to the DNN to introduce a form of conditional learning. Fig. \ref{fig:architecture} depicts the structure of the network as a whole, along with individual components. The network is trained using a simple Binary Cross Entropy loss function, Eq. \ref{eq:bce}, with a Cosine annealed learning rate. Our training strategy consisted of distributing across two Nvidia A30 GPUs with 24GB of VRAM each. We train for a maximum of 100 epochs and deploy an early stopping criterion based on the validation loss. A summary of the training specifications can be found in Table. \ref{tab:training}.

\begin{equation}\label{eq:bce}
    \mathcal{L}_{BCE.} = - \frac{1}{N} \sum_i ^ N \boldsymbol{y}_i \cdot \log(\boldsymbol{\hat{y}}_i) + (1 - \boldsymbol{y}_i) \cdot \log(1 - \boldsymbol{\hat{y}}_i))
\end{equation}

\begin{table}[!]
\setlength{\tabcolsep}{50pt}
\centering
\begin{tabular}{ c  c } 
\hline
\textbf{training parameter} & \textbf{value} \\
\hline
Initial Learning Rate & $\, \, \, 10^{-4}$ \\
Max Epochs & 100 \\
Batch Size & 128 \\
Network memory on local storage & $\sim 82$MB \\
Training Memory per GPU  & $\sim 4GB$ \\
Trainable parameters & 7,084,755\\
Wall Time & $\sim$ 2-3 Days \\
\hline
\end{tabular}
 \caption{\textbf{Training specifications of the classification architecture:} training was performed with two Nvidia A30 24GB GPUs.}\label{tab:training}
\end{table}

The hierarchical structure of a Swin transformer enables the integration of features at different resolutions through patch merging between successive transformer blocks. In downstream Swin blocks, Windowed Multi-Head Self-Attention (W-MHSA) maintains linear complexity relative to the input by computing self-attention within local neighborhoods (windows), avoiding the quadratic complexity of standard vision transformers that compute self-attention across all patches, \textit{i.e.}, traditional MHSA. Sequentially connected layers shift the windows, establishing connections between them \cite{Liu_2021}. 

Eq. \ref{eq:SW_MHSA} illustrates the flow of outputs between two consecutive blocks, $x^k$ and $x^{k+1}$, where $\hat{x}$ and $x$ represent the outputs of the W-MHSA, Shifted Window Multi-Head Self Attention (SW-MHSA), LayerNorm (LN), and Multi-layer Perceptron (MLP) modules \cite{giroux2023}. 

\begin{equation}
    \label{eq:SW_MHSA}
    \begin{aligned}
        \hat{\bold{x}}^l = & \, \text{W-MHSA}(\text{LN}(\bold{x}^{l-1})) + \bold{x}^{l-1} \\
        \bold{x}^l = & \, \text{MLP}(\text{LN}(\hat{\bold{x}}^l)) + \hat{\bold{x}}^l \\
        \hat{\bold{x}}^{l+1} = & \, \text{SW-MHSA}(\text{LN}(\bold{x}^{l})) + \bold{x}^{l} \\
        \bold{x}^{l+1} = & \, \text{MLP}(\text{LN}(\hat{\bold{x}}^{l+1})) + \hat{\bold{x}}^{l+1} 
    \end{aligned}
\end{equation}

In general computer vision tasks, attention mechanisms are thought to teach the network ``where to look''. In our case, we are not necessarily teaching the network where to look, but rather allowing it to find high-information regions in sparse images, and potentially relate them to the general PDF that becomes apparent when integrated over multiple tracks. This is shown in Fig. \ref{fig:sparse_hits}.
We find that the method is powerful at extracting information from the sparse images and requires a significant amount of regularization to prevent overfitting. \footnote{Significant amounts of training data are also needed to provide a good base distribution for the model to learn from.} As such, we utilize Batch Normalization (BN), Dropout and SELU activation functions throughout the network. A description of the hyperparameters used in the network can be found in Table. \ref{tab:params2}.

\begin{table}[!]
\setlength{\tabcolsep}{28pt}
\centering
\begin{tabular}{ c  c  } 
\hline
\textbf{description} & \textbf{value} \\
\hline
Embedding Dimension & 48 \\
Encoder Block Depth & $[2,2,6,2]$ \\
Number of Attention Heads & $[3,6,12,24]$ \\ 
Window Size & 7 \\
Patch Size & $2 \times 2$ \\
Transformer Dropout Probability & $[0.1,0.1,0.1]$ \\
CNN Dropout Probability & $0.2$ \\
\hline
\end{tabular}
 \caption{\textbf{Hyperparameters of the classification architecture:} Hyperparameters for the classification architecture, mainly comprising of parameters for the Swin Transformer. The CNN acts as a lightweight information combination method.}
 \label{tab:params2}
\end{table}

%%%%%%%%%%%%%%%%%%%%%%%%%%%%%%%%%%%%%%%%%%%%%%%%%%%%%%%%%%
\subsection*{Normalizing Flows}
Normalizing Flows (NF) are a method of neural density estimation commonly used when the underlying PDF is unknown. In these cases, it is almost impossible to find analytic representations and therefore likelihood evaluation becomes difficult. NF's aim to curate this through a change of variables, transforming an unknown density into one that is known (such as Gaussian), such that likelihoods can be evaluated analytically. Moreover, we can perform the transformation conditional on a set of parameters such that we can represent the density as a function of the kinematics, $|\vec{p}|,\theta,\phi$. 
Let $\boldsymbol{x} \in \boldsymbol{X}$ denote an element from a set of vectors under an unknown probability distribution $p(\boldsymbol{x}|\boldsymbol{k})$, $\boldsymbol{k} \in \boldsymbol{K}$ the conditional vector for the kinematics of $\boldsymbol{x}$, and $\boldsymbol{z} \in \boldsymbol{Z}$ represent a Gaussian representation of $\boldsymbol{x}$ through a function $f$. \cite{fanelli2022flux+}
A conditional flow with N layers can be described by:
\begin{equation}\label{eq:trans}
\boldsymbol{x}_{\boldsymbol{k}} = f(\boldsymbol{z},\boldsymbol{k}) = f_N \circ f_{N-1} \circ ... f_{1}(\boldsymbol{z_{0}},\boldsymbol{k}),
\end{equation}
where the function $f(\boldsymbol{z},\boldsymbol{k})$ is represented by Affine Coupling transformations \cite{Affine}.

The logarithm of the transformed probability is then given by Eq. \ref{eq:trans_prob}, where $q(\ast |\boldsymbol{k})$ denotes the probability under a Gaussian distribution: 
\begin{equation}\label{eq:trans_prob}
    \log p(\boldsymbol{x}|\boldsymbol{k}) = \log q (f^{-1}(\boldsymbol{x}) | \boldsymbol{k}) - \sum_{i=1}^{N} \log \left|det \left(\frac{\partial f_{i}^{-1}(\boldsymbol{x})}{\partial \boldsymbol{x}} \right)\right|
\end{equation}
The loss function is then given by the negative log-likelihood:

\begin{equation}\label{eq:nf_loss}
    \mathcal{L} = -\frac{1}{|\boldsymbol{X}|}\sum_{\boldsymbol{x},\boldsymbol{k} \in \boldsymbol{X}} \log p(\boldsymbol{x}|\boldsymbol{k})
\end{equation}

The transformed density allows fast simulation through a sampling of the base density, conditional on the kinematics. At training, this looks like a large tabular dataset where each Cherenkov photon associated to a specific track has the same conditional parameters. As a result, we generate individual Cherenkov photons conditional on the kinematics of the track that produced them. The conditional parameters are processed through a NN and produce the mean and covariance matrices of the Multivariate Gaussian distribution. This allows us to remain agnostic to the photon yield and generate proportional to any expected value at a given region of the phase space. We train for a maximum of 300 epochs, using a Cosine Annealed learning rate, and deploy an early stopping criterion based on the validation loss. The hyperparameters of the Flow model can be found in Table. \ref{tab:flow_params}, and training specifications in Table \ref{tab:training_flows}.

\begin{table}[!]
\setlength{\tabcolsep}{28pt}
\centering
\begin{tabular}{ c  c } 
\hline
\textbf{description}  & \textbf{value} \\
\hline
Bijections & 12 \\
Blocks per bijection & 2 \\
Layers per block & 2 \\
Nodes per block layer & 512 \\
Embedding Network Size & [16,16] \\
Block Activation & ReLU \\
\hline
\end{tabular}
 \caption{\textbf{Hyperparameters of the fast simulation architecture:} hyperparameters have been validated through simple grid search methods designed to minimize the performance and complexity tradeoff.
 \label{tab:flow_params}
  }
\end{table}

\begin{table}[!]
\setlength{\tabcolsep}{50pt}
\centering
\begin{tabular}{ c  c } 
\hline
\textbf{training parameter} & \textbf{value} \\
\hline
Initial Learning Rate  & $7\cdot 10^{-4}$ \\
Max Epochs & 300 \\
Batch Size & 2048 \\
Network memory on local storage & $\sim 97$MB \\
Training GPU memory  & $\sim 1GB$ \\
Trainable parameters & 8,438,134\\
Wall Time & $\sim$ 2-3 Days \\
\hline
\end{tabular}
 \caption{\textbf{Training specifications of the fast simulation architecture:} training was performed with a single Nvidia A30 24GB GPU.}\label{tab:training_flows}
\end{table}

It should be clear that in transforming our input distributions from a discrete space in $x,y$, produces generated quantities in a continuous space. However, the discrete, pixelized readout of the DIRC remains and must be accounted for. Moreover, the physical sensor limitations of the DIRC must be taken into account to ensure all generations produced are physical. This can thought of as generating under a prior corresponding to the physical dimensions of the DIRC. To do this we implement a simple resampling technique, programmatically depicted in Algorithm. \ref{code:resampling}.

\begin{algorithm}
\caption{Resample and Update Hits}
\label{code:resampling}
\begin{algorithmic}[1]
    \State \textbf{Input:} Photon yield, Kinematics
    \State \textbf{Output:} Hits
    \State H $\gets$ \Call{GetTrack}{Photon yield, Kinematics}
    \State H$^{\prime}$ $\gets$ \Call{ApplyMask}{H}
    \State $N_{resample.}$ $\gets$ Photon yield - \Call{Length}{H$^{\prime}$}
    
    \While{n\_resample $\neq$ 0}
        \State H$^{\prime\prime}$ $\gets$ \Call{GetTrack}{$N_{resample.}$, Kinematics}
        \State H$^{\prime}$ $\gets$ \Call{Concat}{H$^{\prime}$, H$^{\prime\prime}$}
        \State H$^{\prime}$ $\gets$ \Call{ApplyMask}{H$^{\prime}$}
        \State $N_{resample.}$ $\gets$ Photon yield - \Call{Length}{H$^{\prime}$}
    \EndWhile
    
    \State Hits $\gets$ \Call{SetToClosest}{H$^{\prime}$, physical\_x,physical\_y}
    \State \Return Hits
\end{algorithmic}
\end{algorithm}

For each track, we generate Chereknov photons ($N_{\gamma}$) corresponding to the associated photon yield through our Normalizing Flow. Given the discrete nature of the DIRC, each of the $x,y$ spatial coordinates are rounded. We then apply a masking operation, analogous to a prior, in which we reject unphysical hits in the detector plane across all three dimensions ($x,y$ and time). We then continually resample these photons until the desired photon yield is met. The physical sensor locations are a subset of the discrete set we generate, therefore we perform a translation operation to the nearest allowed pixel.
In doing so, we inherently introduce some bias, although our studies have shown this to be minimal and at maximum at $2.5\%$ effect and is related to the complexity of the phase-space.  This effect (along with the likelihood we learn), is directly correlated to the distributions injected during training. Time information specifically, is very important as the distributions of x and y appear almost uniform on a global scale, as well as strictly bounded. When the density is considered over all three variates, time information is able to break up this uniformity and provide better likelihoods at training time. 

\begin{figure}[!]
    \centering
    \includegraphics[width=0.68\textwidth]{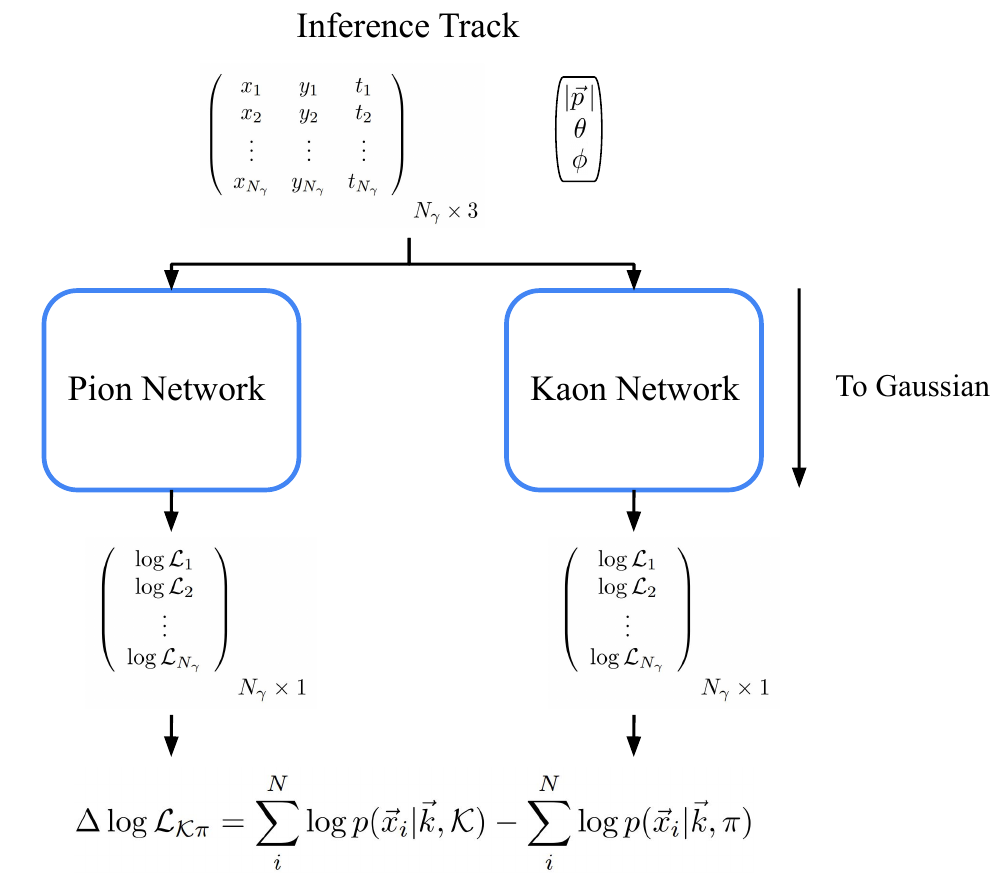}
    \caption{\textbf{Flow chart of Delta-Loglikelihood with two Normalizing Flows:} Individual tracks are represented by matrices of individual Cherenkov photons, conditional on the kinematic parameters. We compute the likelihood of each Cherenkov photon in the base distribution of the normalizing flow for each hypothesis $\pi / K$, such that the total likelihood is the summed contribution of individual hits. The summed quantities are then used to form a DLL on a track-by-track basis. }
    \label{fig:flowchart}
\end{figure}

Moreover, using Eq. \ref{eq:nf_loss}, we can directly compute the likelihood of a point in the base distribution (Gaussian) through an inverse pass by combining the likelihood under a Gaussian to a volumetric correction factor (Jacobian). In our case, this translates to defining a Delta Log-Likelihood (DLL) in the Gaussian space, in which the likelihood of a track is given by the sum of its hits (three-dimensional likelihood in x,y and time). We train two individual networks (one for Kaons, one for Pions) and use these to compute the likelihood of a track under each hypothesis. From this we can form a DLL as in Eq. \ref{eq:DLL}, $\vec{x}_i$ represents an individual Cherenkov photon in a track, $\vec{k}$ represents the kinematics, and $\pi/K$ is the given hypothesis, corresponding to individual networks. A flow chart is depicted in Fig. \ref{fig:flowchart}.

\begin{equation}\label{eq:DLL}
    \Delta \log \mathcal{L}_{K \pi} = \sum_i^N \log p(\vec{x}_i | \vec{k}, K) - \sum_i^N \log p(\vec{x}_i | \vec{k}, \mathcal{\pi})
\end{equation}

%% file: 4_analysis.tex
\section{Analysis and results}\label{sec:results}

\subsection*{Particle Identification}\label{subsec*:PID}

\begin{comment}
We utilize the standard metrics, namely Pion rejection and Kaon efficiency. Associated to these values, is the Area Under the Curve (AUC), in which a perfect PID method will provide an AUC of 1.0. We provide an upper bound of AUC error to be $0.5\%$ (integrated over the phase space) when considering effects of both numerical integration and statistical uncertainty. Fig. \ref{fig:pgun_results} shows the performance of the architecture integrated over the entire phase space (left column), where the AUC is indicated within the legend of each plot. The Swin architecture is shown in the top row, the NF method is shown in the bottom row.

Architecture performance (AUC) as a function of the incoming track momentum (in $500 \; MeV$ bins) is shown in the right column of Fig. \ref{fig:pgun_results}, where the uncertainties are provided through a bootstrapping technique. We estimate the uncertainties on both the rejection and efficiency through Eq. \ref{eq:uncertainty}, where F denotes the desired metric
\end{comment}

\begin{figure}[h!]
    \centering
    \includegraphics[width=0.4\textwidth,height=5.82cm]{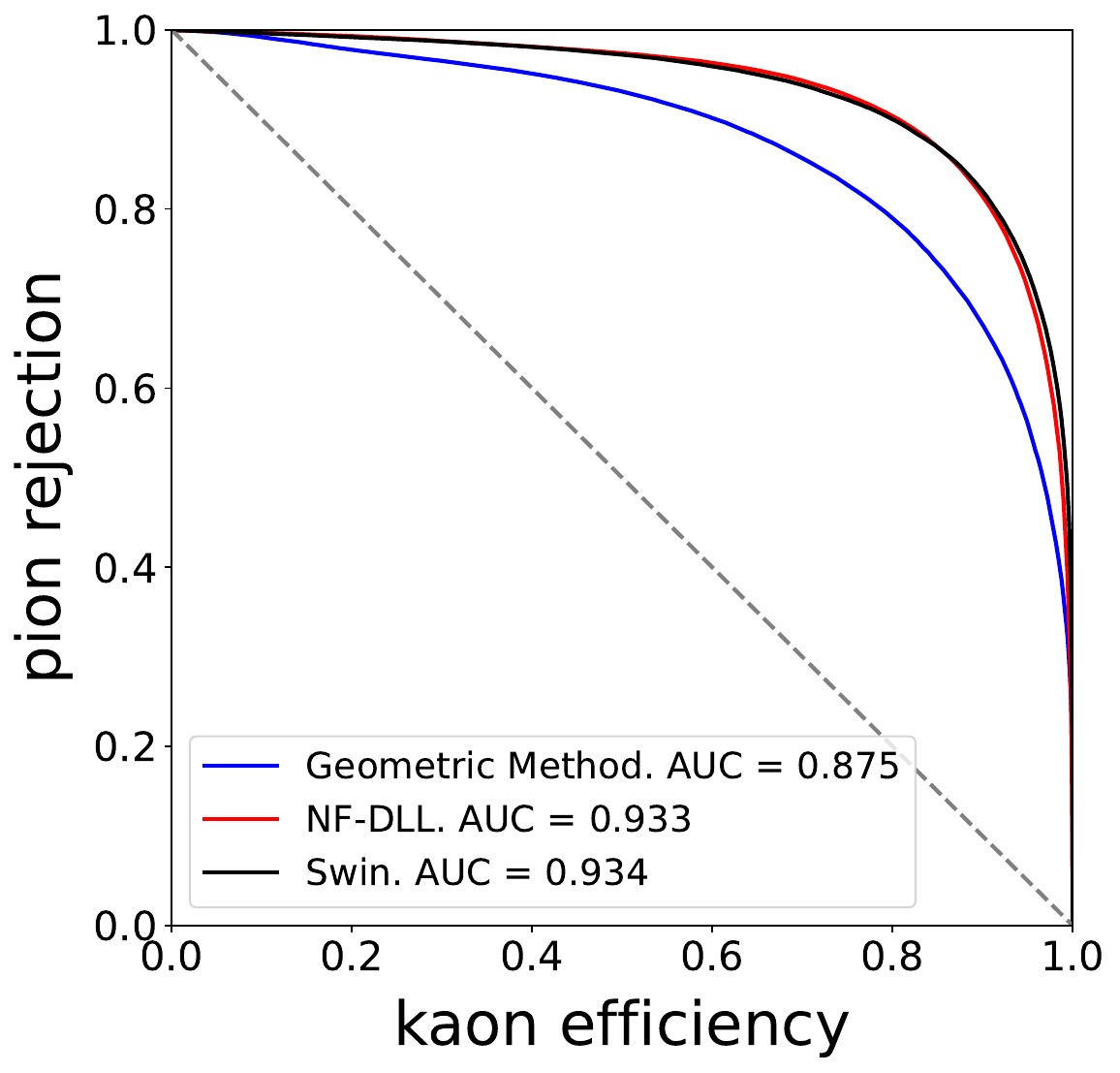} %
    \includegraphics[width=0.485\textwidth,height=5.8cm]{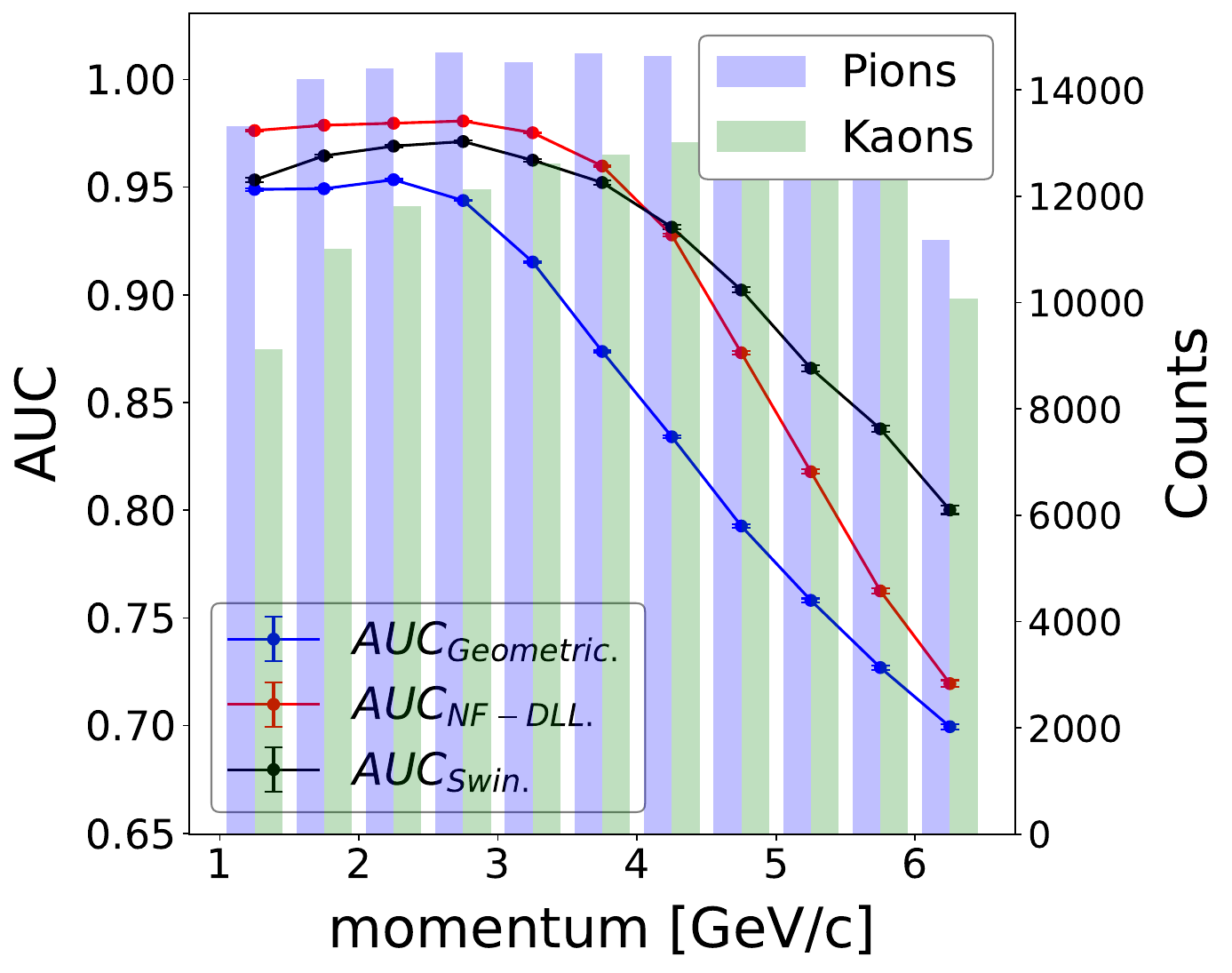}
    \caption{\textbf{Particle gun performance:} Pion rejection as a function of kaon efficiency for the Swin architecture, Normalizing Flow method, compared to the standard geometrical method, integrated over the entire phase space (left). The Area Under the Curve (AUC) is indicated within the legend. AUC as a function of track momentum (right). Uncertainty is represented as the 95\% quantiles obtained through bootstrapping. The counts of pions and kaons 
     for testing is also reported.}
    \label{fig:pgun_results}
\end{figure}

We utilize standard metrics, namely pion rejection and kaon efficiency. Associated with these values is the Area Under the Curve (AUC), where a perfect PID method yields an AUC of 1.0. We provide an upper bound for the AUC error to be $0.5\%$ (integrated over the phase space), considering the effects of both numerical integration and statistical uncertainty. Figure \ref{fig:pgun_results} shows the performance of the architecture integrated over the entire phase space (left column), with the AUC indicated in the legend of each plot. The Swin architecture and the NF method are compared to the established geometrical reconstruction method.
The architecture performance (AUC) as a function of the incoming track momentum (in $500 \; \text{MeV}$ bins) is shown in the right column of Figure \ref{fig:pgun_results}, with uncertainties provided through a bootstrapping technique. We estimate the uncertainties on both the rejection and efficiency using Eq. \ref{eq:uncertainty}, where \( f \) denotes the desired metric, and $N$ denotes the corresponding sample size:
\begin{equation}\label{eq:uncertainty}
    \sigma_f = \sqrt{\frac{f(1-f)}{N}}
\end{equation}

This allows the sampling of different efficiency vs. rejection curves, in which we report the mean AUC value and the associated error as the 95\% quantiles over that specific momentum region. 

From inspection of Fig. \ref{fig:pgun_results} it is apparent our method outperforms the standard geometrical reconstruction when integrated over the phase space, and at all regions of the momentum space. 
Note that AUC is chosen as comparison metric to eliminate the need to apply cuts on the network output. In practice, this cut will be a function of the incoming track momentum due to its relationship with the Cherenkov angle. In this scenario, we plan to use a secondary sub-network to directly regress the ideal threshold. This is left for future optimization.
The Swin Transformer and the DLL method based on normalizing flow (NF-DLL) perform almost equally well up to 4 GeV/c, after which the Swin Transformer outperforms normalizing flow. It should be noted that the DLL method is highly sensitive to noise in the data. With real data and large volumes of data available, we expect the Swin Transformer to be the superior method for classification.

Utilizing an NVIDIA A30 with a batch size of 2048 yields an average inference time of 9 $\mu$s. This performance not only aligns with the previous version of DeepRICH \cite{fanelli2020deeprich}, but the new version, Deep(er)RICH, also expands the PID capabilities of DeepRICH across the entire phase space of the \gluex DIRC detector.

\subsection*{Fast Simulation}

\begin{figure}[h]
    \centering
    \includegraphics[width=0.49\textwidth]{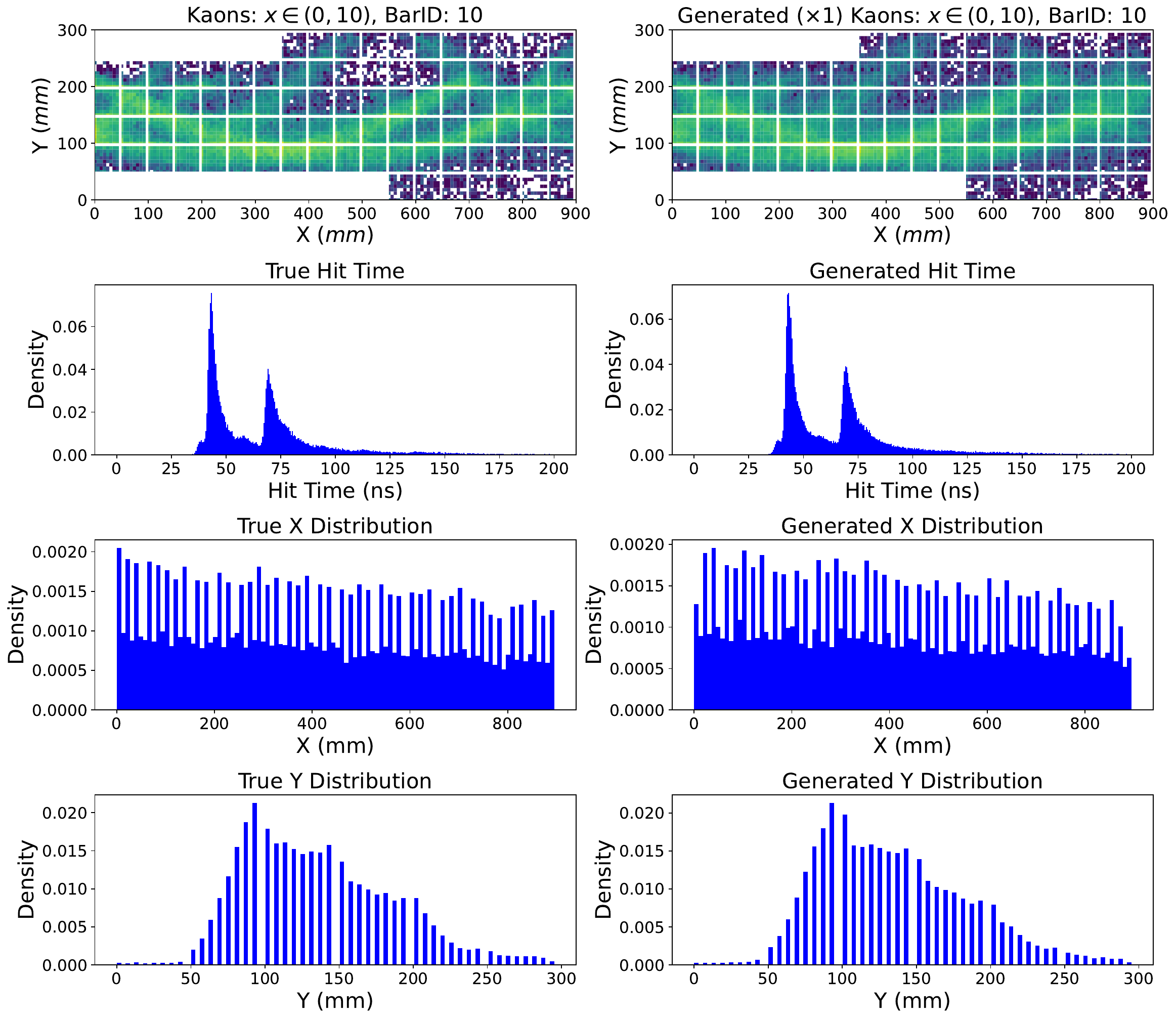} % 
    \includegraphics[width=0.49\textwidth]{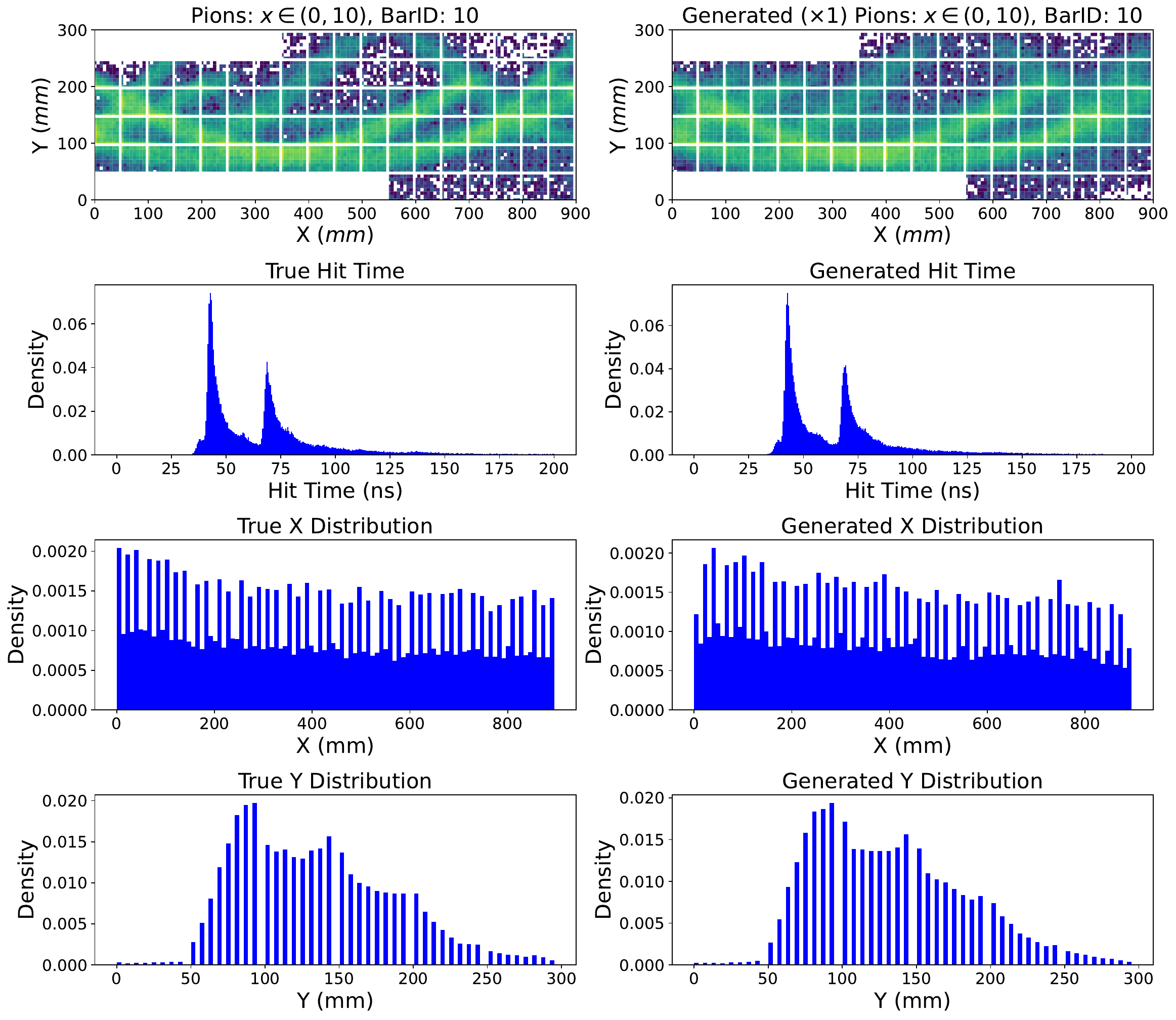}
    \includegraphics[width=0.49\textwidth]{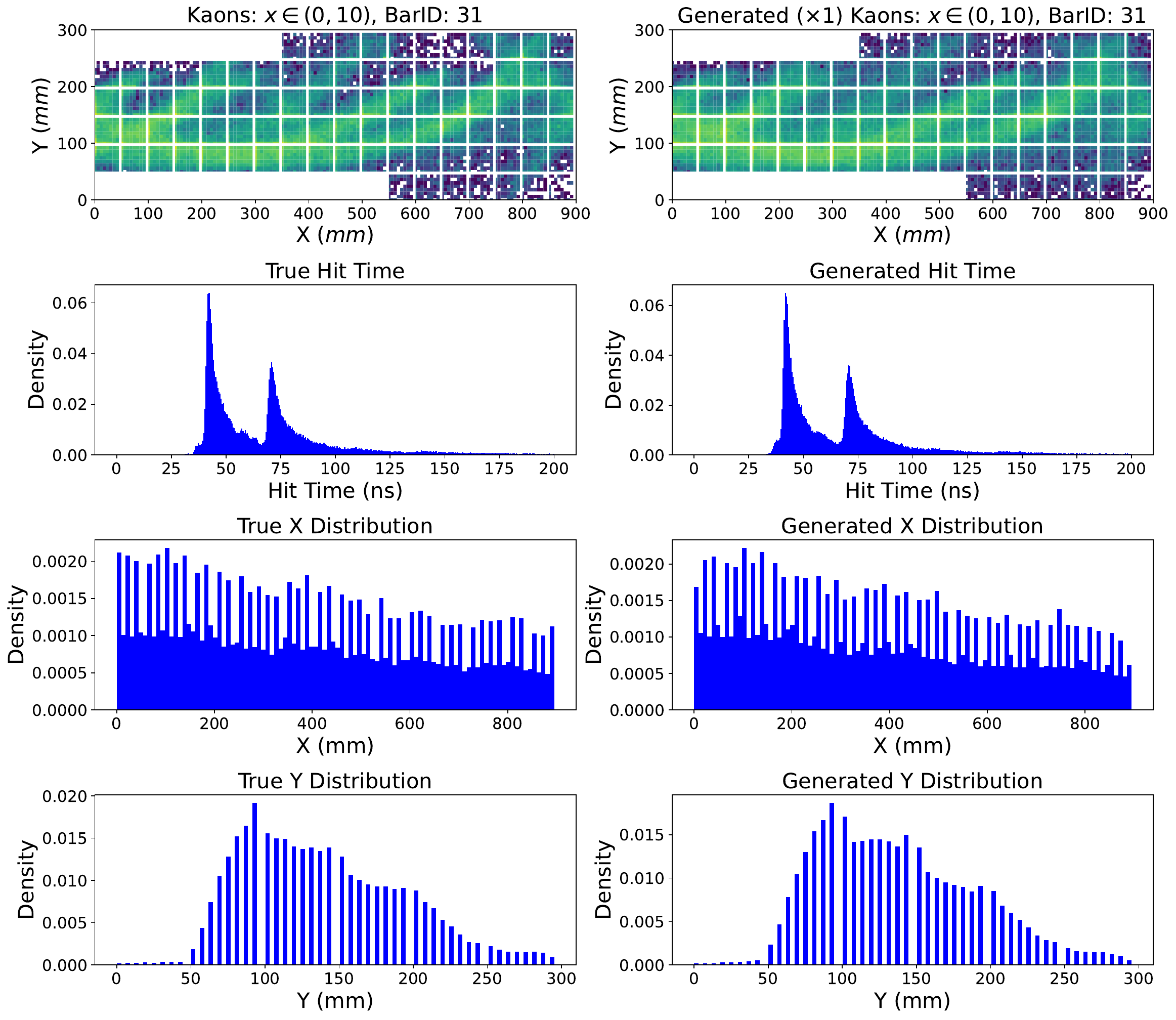} %
    \includegraphics[width=0.49\textwidth]{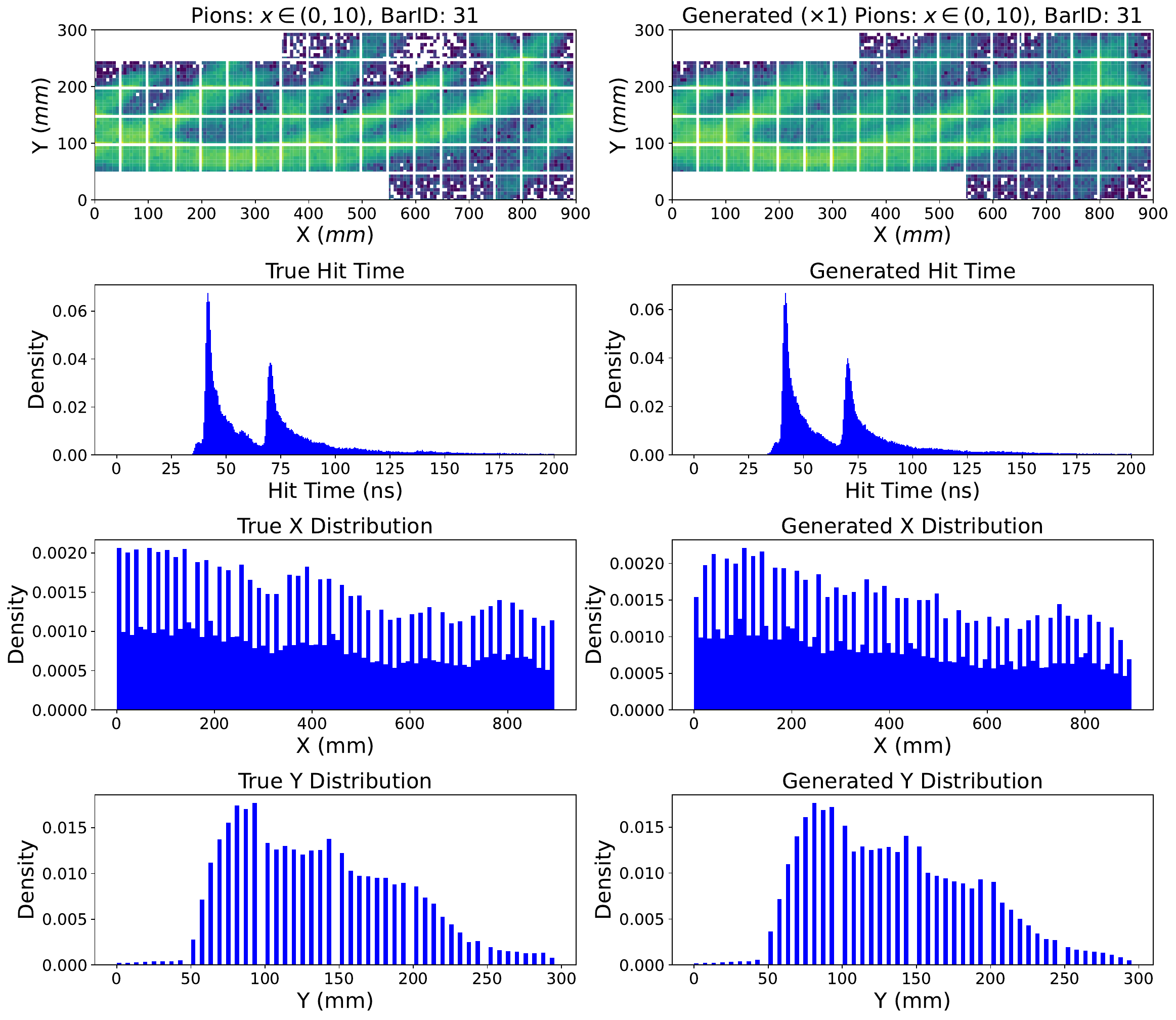}   
    \caption{
    \textbf{Fast Simulation:} Fast simulated samples for pions (left column of two images) and kaons (right column of two images) for \( X \in (0 \,\text{cm}, 10 \,\text{cm}) \) at bar 10 (top row of four images) and bar 31 (bottom row of four images). These regions are centrally located in the DIRC and visualize the two different optical boxes. The fast simulated distributions are shown alongside their ground truth counterparts. Note that in each plot, the photon yield is consistent between the fast simulated and ground truth samples.
}
    \label{fig:generations}
\end{figure}

\begin{figure}[h]
    \centering    \includegraphics[width=0.4\textwidth,height=5.82cm]{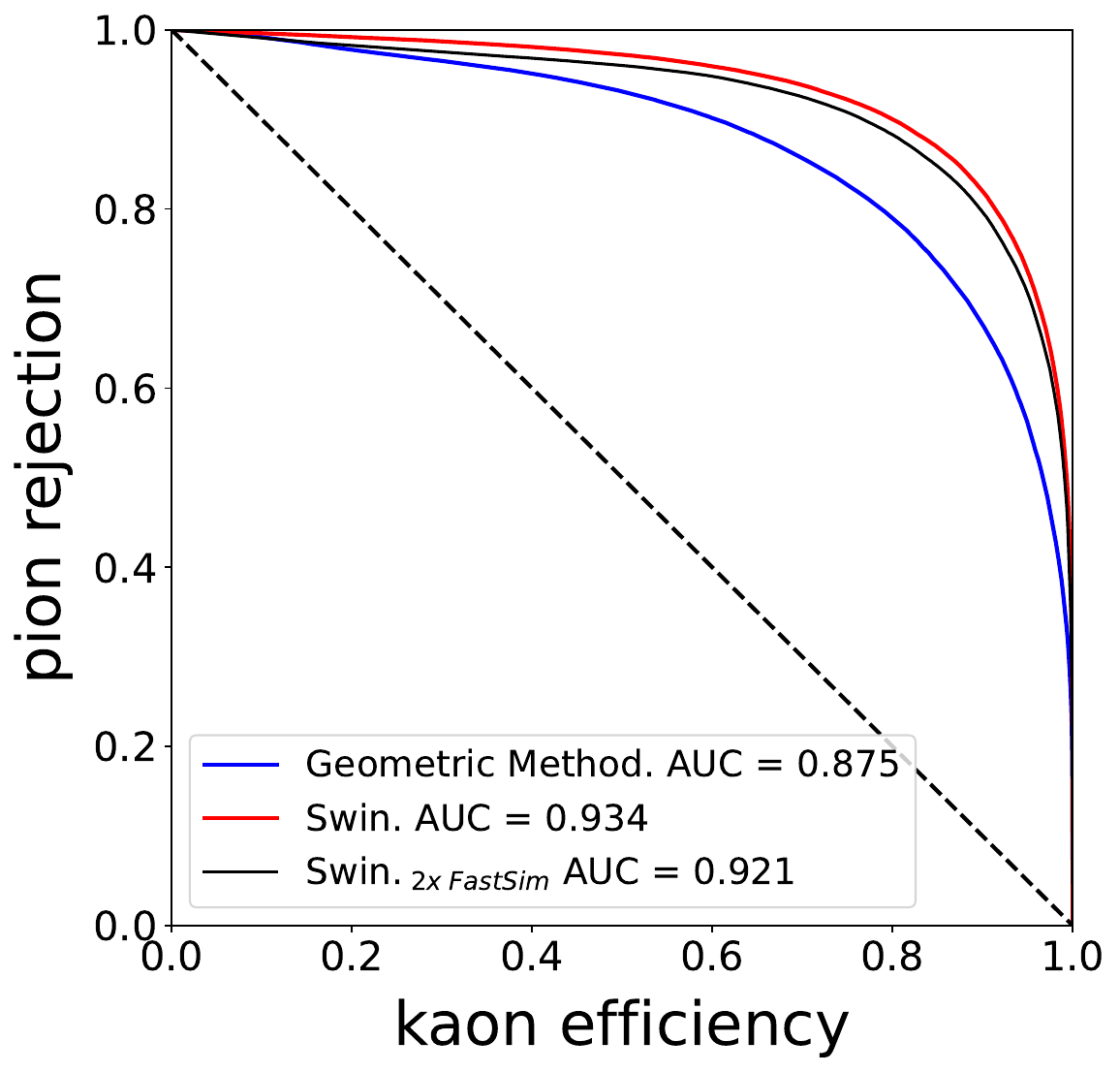} %
    \includegraphics[width=0.485\textwidth,height=5.8cm]{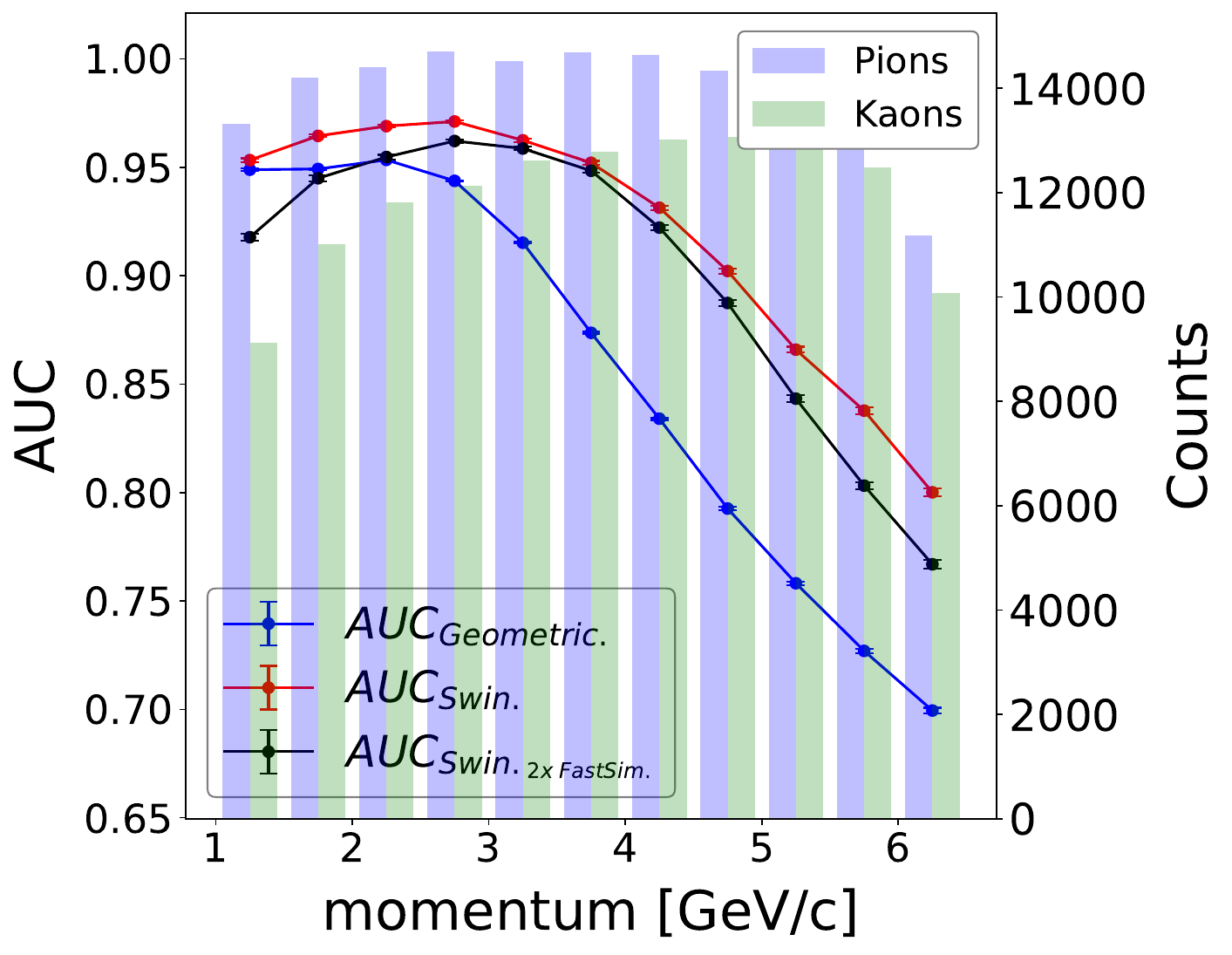}
    \caption{\textbf{Validation of Generations:} Pion rejection as a function of Kaon efficiency (left) for the Swin architecture trained on \geant simulation, the Swin architecture trained on fast simulated data and the standard geometric method, integrated over the entire phase space. The Area Under the Curve (AUC) is indicated within the legend. AUC as a function of track momentum (right). Uncertainty is represented as the 95\% quantiles obtained through bootstrapping}
    \label{fig:fastsim_validation}
\end{figure}

In what follows, we show sample generations from our network in different regions of the phase space. Note that we have produced fast simulations across the entire phase space and have chosen regions with the most complex ring structures for visualization. To do so, we isolate specific bars in the DIRC detector, along with regions in $X$ along the face of the bar. This provides kinematic constraints on the parameters $\theta$ and $\phi$ (within a region) and allows integration over the momentum space to visualize different rings from contributing Cherenkov angles.
Figure \ref{fig:generations} shows pion (left column of two images) and kaon generations (right column of two images) for $X \in (0\,\text{cm}, 10\,\text{cm})$ at bar 10 (top row of 4 images) and bar 31 (bottom row of 4 images). The regions are chosen to be central in the DIRC and provide visualization of the two different optical boxes. The fast-simulated distributions are plotted parallel to their ground truth counterparts. Note that in each plot, between fast-simulated and ground truth, the photon yield is the same.
%
%\( p(\boldsymbol{x}|\boldsymbol{k},\boldsymbol{W}) \), 
%
From visual inspection, it is clear that our model can produce high-fidelity generations of individual Cherenkov photons, retaining the underlying PDF upon integration of multiple tracks. To further validate our fast simulation, we use the classifier developed in the previous section. We train our classifier on fast-simulated pions and kaons and test it on the same Monte Carlo samples used previously. Since we are sampling from the learned probability distribution of our network, we increase the statistics of our training sample to be twice the original training size to ensure sufficient representation of our space. 
We choose the Swin Transformer classifier as a method of validating the similarity between fast simulated and original data, over standard metrics such as Maximum Mean Discrepancy (MMD) or Wasserstein Distance, due to complications arising from interpreting their values in an absolute sense.
Figure \ref{fig:fastsim_validation} shows a comparative performance of the Swin architecture trained on fast-simulated data and the original Monte Carlo sample. The ROC curve integrated over the phase space is shown on the left, while the AUC as a function of momentum is shown on the right.
As mentioned previously, we quote an upper bound error on the AUC integrated over the phase space of $0.5\%$, indicating agreement between the models within statistical uncertainty. The classifier exhibits equivalent performance whether trained on original datasets or those generated by the proposed normalizing-flow-based fast simulation method, with consistent statistical volumes, demonstrating its inability to distinguish fast simulated data from original data. Additionally, we illustrate how our fast simulation can augment the original dataset, enhancing the performance of the data-hungry transformer model.
The AUC as a function of momentum plot shows some disagreement when statistical uncertainty is considered in select regions of the phase space, but overall, the agreement is reasonable. This disagreement is likely to be reduced through an increased training sample size for the fast simulation networks, given that the conditional parameters are continuous.

Previous efforts such as FastDIRC \cite{hardin2016fastdirc} have successfully replicated PMT images from full \geant-based Monte Carlo simulations, achieving this with over 10,000 times less CPU time. The simulation time is approximately 300 ms for a single particle, generating $\mathcal{O}(10^{4}-10^{5})$ hits, which translates to an effective time per hit of $\mathcal{O}(3-30) \mu s$. In contrast, our fast simulation architecture with Deep(er)RICH, optimized for GPU deployment, can generate $\mathcal{O}(3.5 \cdot 10^{6})$ hits in a batch at inference in about 2.5 s on an NVIDIA A30 GPU, yielding an effective time per hit of $\mathcal{O}(0.5) \mu s$. It is worth highlighting a major difference between methods like FastDIRC and Deep(er)RICH. FastDIRC is a fast simulation method that utilizes clever approximations to produce hit patterns and can be tuned to reproduce the expected hit patterns obtained with accurate toolkits such as \geant. In contrast, Deep(er)RICH directly learns how to reproduce the hit patterns obtained by \geant or directly measured from real data.

%% file: 5_impacts.tex
\section{Physics applications and broader impacts}\label{sec:impacts}

The new methods introduced for near real-time PID with Swin Transformers and fast simulations with Normalizing Flows for imaging Cherenkov detectors have multiple physics applications and a broad impact, affecting experiments at JLab and the ePIC experiment at EIC, along with other experiments utilizing imaging Cherenkov detectors.
Notably, in this paper, we demonstrated applications for a specific type of Cherenkov detector, the DIRC, which is operating in \gluex. This approach can be easily extended to the future high-performance DIRC that will cover the barrel region of the ePIC detector \cite{kalicy2024high}. 
Deep(er)RICH can be also utilized for other types of imaging Cherenkov detectors, such as RICH detectors. Examples include the dual-Radiation Imaging Cherenkov (RICH) system in the hadron endcap \cite{cisbani2020ai} and the proximity focus RICH in the electron endcap \cite{khalek2022science} for the ePIC detector.
Our method can be readily adapted to any experiment, simply by retraining on different datasets.
In the following, we describe a selection of potential applications for PID and fast simulations and the impact of their combination.

%%%%%%%%%%%%%% Transformer 

\subsection*{Particle Identification}

The effective inference time per particle of 9 $\mu s$ indicates the potential for near real-time PID applications with the DIRC. 
Fast deep learning architectures can be integrated into high-level trigger systems, which aim to accommodate increased luminosity using dedicated GPUs for data processing. They could also be integrated into the streaming readout schemes currently being designed for the \epic experiment \cite{bernauer2023scientific}.\footnote{It is worth reminding that the expected physics rate for electron-proton collisions in \epic, excluding background, is approximately 500 kHz, with data rates for the hpDIRC in the barrel region notably lower than those in the endcap regions.}

The DIRC detector significantly impacts the strangeness physics program of \gluex \cite{pauli2022strangeness}. It can also enhance the near-threshold $J/\psi$ physics program at high luminosity, critical for understanding the proton's gluon structure, mass radius, the trace anomaly's contribution to proton mass, and the existence of possible hidden-charm pentaquarks \cite{lhcb2019observation, aaij2015observation, winney2023dynamics, winney2019double}. \gluex's results indicate complex structures in the total cross-section energy dependence and suggest contributions beyond simple gluon exchange in the differential cross-section near the threshold, consistent with the hypothesized roles of open-charm intermediate states. A typical $\Lambda_c^{+} \bar{D}^{*}$ event signature, featuring five charged tracks (\textit{e.g.}, $p$, $K^{-}$, $\pi^{+}$, $K^{+}$, $\pi^{-}$) and one or two photons, underscores the need for efficient tracking and PID. Many such events involve at least two charged tracks within the DIRC, necessitating its robust performance as a function of the particle kinematics.

%%%%%%%%%%% Fast Simulation 

\subsection*{Fast Simulation}

Cherenkov detectors require the simulation of optical processes involving numerous photons interacting with complex optical elements, a computation-intensive task typically performed using \geant. Our effective time per hit of $\mathcal{O}(0.5) \mu s$ represents the state of the art for fast simulations with DIRC detectors.
Our work extends beyond the methodologies described in \cite{derkach2020cherenkov}, where a Generative Adversarial Network is trained to reproduce high-level features, specifically the likelihood results from FastDIRC \cite{hardin2016fastdirc}. The distinctive feature of our architecture is its ability to learn the detector response directly from real data, a capability unique to Deep(er)RICH. This makes it stand out compared to other classical fast simulation methods that need to be tuned to match real data. Our approach learns to simulate the detector response at the hit level based on kinematics, offering a more comprehensive simulation strategy. Consequently, a key advantage of our architecture is its superior portability compared to methods like FastDIRC, which require detailed geometric information of the experiment.
Our new method enhances the fast simulation of imaging Cherenkov detectors by extending the applicability across the entire phase space of the detected charged particles, achieving high-fidelity simulations across the full kinematics covered by these detectors.
This advancement builds upon our previous DeepRICH method \cite{Fanelli_2020}, offering a more accurate simulation of the original distributions. Specifically, the integration of normalizing flows yields superior accuracy compared to the variational autoencoders used in the earlier DeepRICH approach.
Another obvious consequence of using high-fidelity fast simulations trained on high-purity samples from real data is that the simulated hit patterns would be inherently aligned and calibrated with real data.

\subsection*{Combining enhanced PID and Fast Simulation}
Combining enhanced PID and fast simulations, Deep(er)RICH allows to manage complex topologies involving multiple tracks simultaneously detected in the same optical box of a DIRC detector. 
For example, the \epic hpDIRC detector, featuring twelve optically isolated sectors arranged in a 12-sided polygonal barrel geometry \cite{kalicy2024high} with 4900 mm long fused silica bars, is built to provide charged particle identification over the kinematics of interest for ePIC. This capability is critical for the ePIC physics program, particularly in analyzing Semi-Inclusive Deep Inelastic Scattering (SIDIS) events \cite{seidl2023ecce}.
Preliminary analyses using an 18 GeV electron beam and a 275 GeV proton beam with loose SIDIS criteria reveal that over 10\% of these events involve at least two charged tracks with momenta above 1 GeV/c detected simultaneously in one sector of the DIRC \cite{cpecar}. Our method allows to overlay multiple hit patterns each originating from one of the simultaneously detected charged particles, assessing if their combined hit pattern matches the observed topology. Unlike established methods, which struggle with overlapping tracks with patterns detected in the same optical box, our approach promises substantial advancements.

%% file: 6_summary.tex
\section{Summary and conclusions}\label{sec:summary}

The Deep(er)RICH architecture developed in this paper, combining Swin Transformers and normalizing flows, shows promising results for both particle identification—extending previous DeepRICH results to the entire kinematic region covered by the DIRC detector in \gluex—and for fast simulation, achieving the first realization of complex hit pattern simulations by directly learning from data. 
Deep(er)RICH learns these tasks continuously as a function of the charged particle kinematics, characterized by momentum and direction, along with the point of impact in the DIRC plane.
We showed the high quality and stability of the reconstruction within the kinematic region and demonstrated superior performance compared to established methods such as the geometrical reconstruction method for PID.

We demonstrated the high fidelity of our fast simulations supported by normalizing flows. A closure test based on transformer classification showed identical performance when trained on fast-simulated or original data, indicating the inability to distinguish between their hit patterns.
Leveraging GPU deployment, we have achieved state-of-the-art time performance, with an effective inference time for identifying a charged particle of 9 $\mu s$, comparable to the first version of DeepRICH, and an effective simulation time of 0.5 $\mu s$ per hit. This enables near real-time applications, which are of particular interest for future high-luminosity experiments aiming to implement deep learning architectures in high-level triggers or more sophisticated streaming readout schemes like those under development at the EIC.
The high quality of reconstruction and the fast computing time are two compelling features of the Deep(er)RICH architecture. The possibility of combining enhanced PID and fast simulations also enables handling complicated topologies arising from overlapping hit patterns detected in the same optical box and generated by simultaneously detected tracks, a problem that traditional methods currently cannot cope with. Consequently, Deep(er)RICH could contribute to important physics channels at both JLab and EIC, as discussed in this paper.
%j
Deep(er)RICH is extremely portable; it is agnostic to the data injected and the detector geometry, and can therefore be adapted to other experiments and imaging Cherenkov detectors beyond the DIRC. 
Deep(er)RICH has also been designed to be easily generalized to classify other categories of particles beyond $\pi/K$, with extending the network left for future development.
Another suggestive application could be training Deep(er)RICH using high purity samples of particles from real data, allowing it to deeply learn the response of Cherenkov detector.